# Structural and magnetic properties of hexagonal perovskites $La_{1.2}Sr_{2.7}MO_{7+\delta}$ ($M$ = Ru, Ir)

**Thomas Götzfried, Armin Reller and Stefan G. Ebbinghaus**

*Lehrstuhl für Festkörperchemie, Institut für Physik, Universität Augsburg, 86159 Augsburg, Germany*

The structures of the new compound $La_{1.2}Sr_{2.7}IrO_{7+\delta}$ and the recently discovered $La_{1.2}Sr_{2.7}RuO_{7+\delta}$ have been solved using a combination of X-ray and neutron diffraction. Both compounds crystallize in the trigonal space group $R$-$3m$ and contain isolated $MO_6$ ($M$ = Ru, Ir) octahedra, which are arranged in well-defined layers. Results of the magnetic susceptibility and XANES measurements show that the transition metal cations are in a pentavalent state. While in $La_{1.2}Sr_{2.7}RuO_{7+\delta}$ an antiferromagnetic interaction between the $Ru^{5+}$ ions is found, $La_{1.2}Sr_{2.7}IrO_{7+\delta}$ shows a very small temperature-independent paramagnetism down to 1.8 K due to the strong spin-orbit coupling characteristic for the 5$d$ element iridium.

## Introduction

The solid-state chemistry of perovskite-related oxides containing 4$d$ and 5$d$ transition metals has attracted a great deal of attention during the last two decades. Besides a wide range of magnetic and electronic properties, these materials show a remarkable compositional flexibility. Even though a large variety of compositions and structure types have been identified, they primarily involve alternate stacking of perovskite layers with layers of other structural types. Ruddlesden-Popper compounds $A_{n+1}B_nO_{3n+1}$[1,2] or $A_nB_nO_{3n+2}$ phases (e.g. $Sr_2Nb_2O_7$, $n = 4$[3,4]) are well-known examples consisting of perovskite slabs, the thickness of which is described by the value $n$. These families of compounds have been the subject of many investigations due to their low-dimensional electrical and magnetic properties.

In the class of hexagonal perovskites, the low-dimensional character is even more pronounced. These oxides consist of chains of face-sharing $BO_6$ octahedra separated by the A cations. Due to the short distance of the B cations (approximately 2.5 Å); direct intermetallic interactions can occur. The $BO_6$ chains can be interrupted by e.g. trigonal prisms containing diamagnetic ions. This results in discrete dimers, trimers etc. of paramagnetic centers, in which interesting coupling phenomena are found.[5,6]

Our research on hexagonal perovskites focuses on compounds containing platinum group metals in unusually high oxidation states, e.g. pentavalent ruthenium or iridium ions. Such highly oxidized cations from the second or third transition series[7] often show quite unusual magnetic properties.

In this paper, we describe two compounds belonging to the $(A'_2O_{1+\delta})(A_nB_{n-1}O_{3n})$ series of hexagonal perovskite oxides. Very few oxides belonging to this family have been reported to date: one example is the system $Ba_5Ru_2O_{10}/Ba_5Ru_2O_9(O_2)$[8,9] corresponding to $n = 3$ with $\delta = 0$ and 1, respectively. For $\delta = 1$, the two oxygens in the $(A'_2O_{1+\delta})$ layers form peroxide units $(O_2^{2-})$. This particular feature is also found in the system $Ln_2Ca_2MnO_{7+\delta}$,[10-12] an $n = 2$ member of this series. In a recent article, we reported on the new 4$d$ transition metal oxide $La_{1.2}Sr_{2.7}RuO_{7+\delta}$.[13] Structural investigations based on X-ray diffraction data showed $La_{1.2}Sr_{2.7}RuO_{7+\delta}$ to be isostructural to $La_2Ca_2MnO_7$. Figure 1 (left) represents the $n = 2$ crystal structure viewed along the (110) direction. The structure can be described by the stacking of close-packed $(AO_3)$ layers, according to the sequence αββγγα…, divided by one intermediate $(A'_2O_{1+\delta})$ layer (i) to give the stacking sequence αβiβγiγα… The transition metal cations occupy the octahedral interstitial sites between neighboring $(AO_3)$ layers resulting in the formation of hexagonal perovskite-related $A_2BO_6$ slabs. In this arrangement the B-type cations form isolated octahedra separated by $(A'_2O_{1+\delta})$ layers along the $c$-axis. Figure 1 (right) illustrates one of these octahedron layers reflecting the trigonal symmetry. XRD-Rietveld refinements together with thermogravimetric measurements led to a value of $\delta \approx 1/3$ and revealed the presence of both oxide and peroxide ions in the $(A'_2O_{1+\delta})$ layers. To gain a more detailed insight into the oxygen coordination especially in these layers, we here present structure refinements using neutron diffraction data.

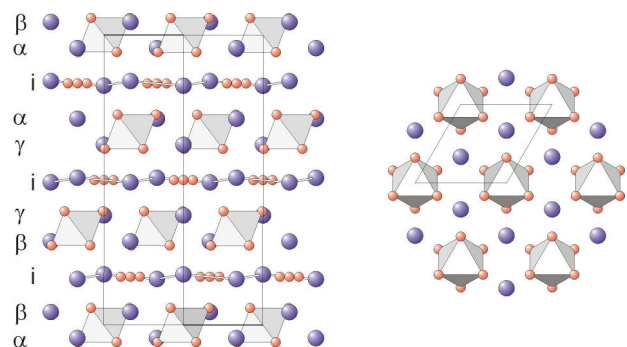

**Fig. 1** Left: Projection of the crystal structure of $La_{1.2}Sr_{2.7}MO_{7+\delta}$ ($M$ = Ru, Ir) along the (110) direction; greek letters represent the $(AO_3)$ layers (A = La/Sr), i the intermediate $(A'_2O_{1+\delta})$ layers (A' = Sr). The transition metal cations lie within the octahedra. Right: Hexagonal perovskite layer $(La,Sr)_2MO_6$ ($M$ = Ru, Ir) viewed down the $c$-direction.

In a systematic study to replace ruthenium by other noble metals we managed to prepare the new compound $La_{1.2}Sr_{2.7}IrO_{7+\delta}$ which turned out to be the Ir-analogue of the above-mentioned ruthenate. $La_{1.2}Sr_{2.7}IrO_{7+\delta}$ is the first $n = 2$ oxide of the $(A'_2O_{1+\delta})(A_nB_{n-1}O_{3n})$ series containing a 5$d$ transition metal. Here we report structural and magnetic properties of the iridate determined by combined X-ray and neutron powder diffraction and by measurements of the magnetic susceptibility and X-ray absorption spectroscopy.

## Experimental

Polycrystalline samples of $La_{1.2}Sr_{2.7}MO_{7+\delta}$ ($M$ = Ru, Ir) were prepared by heating appropriate amounts of decarbonated $La_2O_3$, $SrCO_3$, $RuO_2$, and Ir metal, respectively, in air at 1250°C for 36 h. The products were slowly cooled to 900°C within 35 h, after which the furnace was switched off. The obtained powders were of reddish-brown and black color for the ruthenate and iridate, respectively.

X-ray powder diffraction measurements were carried out on a Seifert 3003-TT diffractometer in the range of $13° \leq 2\theta \leq 120°$ with a step width of 0.015° and a measuring time of 10 s per data point using Cu-$K_\alpha$ radiation.

The neutron powder diffraction patterns were recorded on the high-resolution powder diffractometer for thermal neutrons (HRPT)[14] at PSI, Villigen (Switzerland). The samples (about 3 g) were enclosed in a cylindrical vanadium container of 8 mm inner diameter. Data collection was performed at room temperature with the following experimental conditions: $\lambda = 1.494$ Å, $2\theta$ range 5-165°, $2\theta$ step size 0.05°.

A simultaneous Rietveld refinement of the XRD and ND data was performed with FullProf2k Multi-Pattern.[15] For the neutron data, a Thompson-Cox-Hastings pseudo-Voigt peak-shape function was used, whereas the X-ray data were fitted with a pseudo-Voigt function. The background of each data set was described by interpolation between a set of fixed points. The neutron patterns, which were recorded in transmission mode, were absorption-corrected. The calculated absorption correction coefficient ($\mu R$) was found to be 0.1 for $La_{1.2}Sr_{2.7}RuO_{7+\delta}$, whereas for the iridium containing compound a value of 2.0 was obtained. Peaks caused by the vanadium container contributed only marginally to the diffraction pattern and a trace impurity of $La(OH)_3$ was included in the refinements of the X-ray diffraction data. X-ray quantitative analysis was used to establish the volume fraction of this phase. It turned out that the fraction was well below 1% for the title compounds.

The temperature dependence of the magnetic susceptibility was investigated using a Quantum Design MPMS SQUID magnetometer. Samples were measured under both zero-field cooled (ZFC) and field cooled (FC) conditions in the temperature range $1.8\ K \leq T \leq 300\ K$.

X-ray absorption experiments at the Ir-$L_{III}$ edge were carried out at the beamline A1 of HASYLAB at DESY. XANES spectra of the title compound and several reference materials were recorded in transmission mode at room temperature. Appropriate amounts of the samples were mixed with polyethylene and pressed into pellets to obtain an edge step $\Delta\mu d \sim$ 1-2. Reproducibility of the determination of the edge position was achieved by recording spectra of a Pt metal foil before and after each measurement of a series of compounds. The obtained raw data were energy calibrated, background corrected (Victoreen fit) and normalized.

## Results and Discussion

Structure refinement of $La_{1.2}Sr_{2.7}RuO_{7+\delta}$

A structural model for this compound achieved by XRD Rietveld refinement has been reported recently.[13] In this paper, we present a joint refinement of X-ray and neutron diffraction data that provides a much deeper insight in some structural aspects. Due to the high neutron scattering length of oxygen, detailed information about the oxygen stoichiometry as well as metal–O bond lengths, angles and (anisotropic) displacement parameters can be derived. Refinement results are given in Table 1, and selected interatomic distances are listed in Table 2. Figure 2 shows the profile fits to the neutron and X-ray diffraction data, respectively.

The most interesting structural feature of the title compounds is the oxygen position within the ($A'_2O_{1+\delta}$) layers. The oxygen ions do not occupy the $3b$ site, i.e. the center of the hexagonal cavity formed by the A ions, but are delocalized on an off-center position. In previous reports,[11,12,16] refinements were carried out with the oxygen occupying the $18g$ site. On the basis of X-ray and neutron diffraction data, difference Fourier analysis with the O2 ion removed from the model clearly showed not only residual scattering density at the $18g$ site but a rather donut-shaped residual density. Therefore, the oxygen ions were allowed to occupy both the six-fold degenerate $18g$ site and the intermediate positions ($18h$ site). Identical displacement parameters were used for these sites, whereas the occupancies were refined independently. It turned out that the $18g$ site is indeed preferably occupied consistent with previous results. From the SOFs, a total oxygen content of 1.31(3) is obtained for the O2 site. This occupancy is in very good agreement with findings based on thermogravimetric and X-ray Rietveld analysis reported recently.[13] A value greater than unity implies that part of the oxide ions are replaced by peroxide ions. The occurrence as both oxide and peroxide is characteristic for this family of compounds. It is noteworthy that the replacement of $O^{2-}$ (oxide ion) by $(O_2)^{2-}$ (peroxide ion) does not change the oxidation state of the transition metal ion. For the pure oxide systems it was found that the $O^{2-}$ ions only occupy the $18g$ site.[12,16] Provided that this is also valid for mixed oxide/peroxide systems, the peroxide ions might only be located at the $18h$ position. However, the $18g:18h$ ratio was found to be approximately 3:1 ruling out this separate ion arrangement with $O^{2-}$ on the $18g$ and $(O_2)^{2-}$ on the $18h$ sites. We therefore conclude that the peroxide ions occupy one 18$g$ and one 18$h$ site. This is supported by the fact that the bond distance of 1.533(28) Å (Figure 3) is comparable to typical bond lengths of peroxide ions observed in the previously synthesized compounds $Ba_5Ru_2O_{11}$ (1.554 Å)$^9$ and $BaO_2$ (1.49 Å)$^{17}$, respectively. This setting leads to an oxide to peroxide ratio of about two to one, i.e. the $\delta$ value is about one third as expected from the SOFs. Furthermore, it should be noted that the distance of 1.595 Å between two 18$h$ positions is also in the range of peroxide bond lengths (see Figure 3).

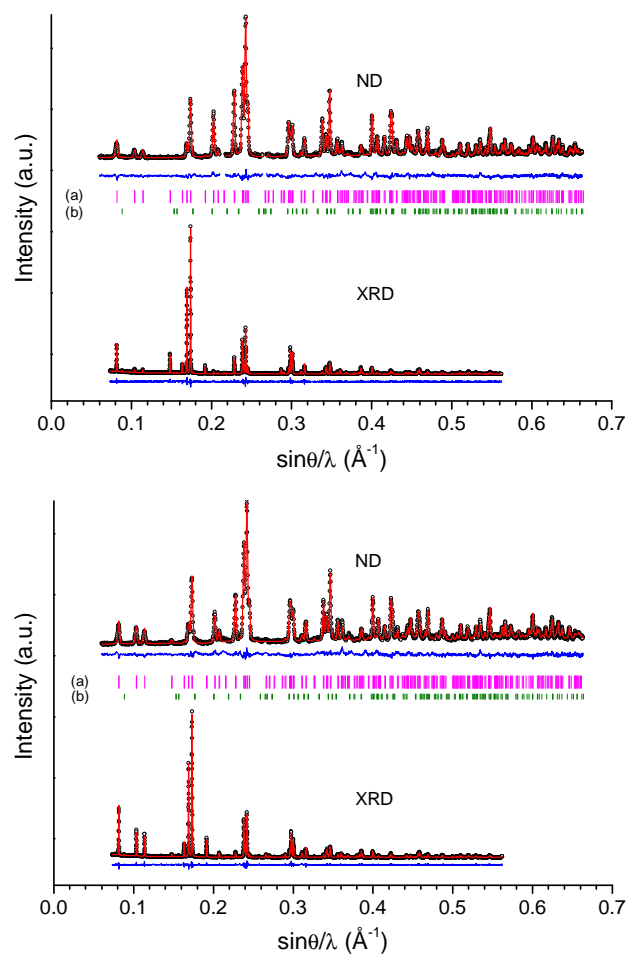

**Fig. 2** Observed (open circles), calculated (solid line) and difference (bottom) X-ray and neutron powder diffraction patterns for $La_{1.2}Sr_{2.7}RuO_{7+\delta}$ (top) and $La_{1.2}Sr_{2.7}IrO_{7+\delta}$ (bottom). Bragg reflections for the respective compound (a) and $La(OH)_3$ (b) are represented by tick marks. The insets show the most intense peaks of the neutron data.

**Table 1** Atomic coordinates refined with X-ray and neutron diffraction data of $La_{1.2}Sr_{2.7}MO_{7+\delta}$ ($M$ = Ru, Ir) at room temperature

| Atom | Site | SOF | $x$ | $y$ | $z$ | $B_{eq}$ (Å$^2$) |
|---|---|---|---|---|---|---|
| $La_{1.2}Sr_{2.7}RuO_{7+\delta}$ | | | | | | |
| $a$ = 5.753(1) Å, $c$ = 18.351(3) Å | | | | | | |
| X-ray data: $R_p$ = 0.0820, $R_{wp}$ = 0.110 | | | | | | |
| neutron data: $R_p$ = 0.0308, $R_{wp}$ = 0.0399 | | | | | | |
| Ru | 3$a$ | 1 | 0 | 0 | 0 | 0.619(37) |
| La1/Sr1 | 6$c$ | 0.578(15)/0.422(15) | 0 | 0 | 0.62326(7) | 0.904(30) |
| Sr2 | 6$c$ | 0.709(20) | 0 | 0 | 0.1753(2) | 0.72(8) |
| Sr2a | 18$h$ | 0.065(6) | 0.099(5) | 0.050(5) | 0.1744(9) | 0.72(8) |
| O1 | 18$h$ | 1 | 0.3157(4) | 0.1576(4) | 0.0611(1) | 3.562$^a$ |
| O2a | 18$g$ | 0.158(10) | 0.1155(15) | 0 | 0.5 | 0.74(18) |
| O2b | 18$h$ | 0.061(9) | 0.092(9) | 0.185(9) | 0.5 | 0.74(18) |
| $La_{1.2}Sr_{2.7}IrO_{7+\delta}$ | | | | | | |
| $a$ = 5.771(1) Å, $c$ = 18.348(3) Å | | | | | | |
| X-ray data: $R_p$ = 0.0820, $R_{wp}$ = 0.112 | | | | | | |
| neutron data: $R_p$ = 0.0338, $R_{wp}$ = 0.0438 | | | | | | |
| Ir | 3$a$ | 1 | 0 | 0 | 0 | 0.767(29) |
| La1/Sr1 | 6$c$ | 0.557(12)/0.443(12) | 0 | 0 | 0.62293(7) | 0.947(36) |
| Sr2 | 6$c$ | 0.788(16) | 0 | 0 | 0.1753(1) | 1.03(9) |
| Sr2a | 18$h$ | 0.047(5) | 0.123(7) | 0.062(7) | 0.1717(14) | 1.03(9) |
| O1 | 18$h$ | 1 | 0.3194(5) | 0.1595(5) | 0.0616(1) | 3.581$^b$ |
| O2a | 18$g$ | 0.165(11) | 0.1157(19) | 0 | 0.5 | 0.80(21) |
| O2b | 18$h$ | 0.064(11) | 0.093(10) | 0.19(10) | 0.5 | 0.80(21) |

$^{a,b}$ Refined with anisotropic displacement parameters (Å$^2$):
$^a$ $\beta_{11}=2\beta_{12}$, $\beta_{22}$=0.0654(10), $\beta_{33}$=0.0011(1), $\beta_{12}$=0.0039(7), $\beta_{13}=2\beta_{23}$, $\beta_{23}$=-0.0001(1)
$^b$ $\beta_{11}=2\beta_{12}$, $\beta_{22}$=0.0625(11), $\beta_{33}$=0.0013(1), $\beta_{12}$=0.0046(8), $\beta_{13}=2\beta_{23}$, $\beta_{23}$=-0.0002(2)

The displacement ellipsoids for the oxygen ions on the O1 site were found to be rather elongated (Figure 4), indicating that either the anions are thermally mobile or possess a considerable positional disorder. To distinguish

**Table 2** Selected bond lengths (Å)

| | $La_{1.2}Sr_{2.7}RuO_{7+\delta}$ | $La_{1.2}Sr_{2.7}IrO_{7+\delta}$ |
|---|---|---|
| 6 × Ru-O1 | 1.932(2) | - |
| 6 × Ir-O1 | - | 1.957(3) |
| 3 × La1/Sr1-O1 | 2.596(2) | 2.599(3) |
| 6 × La1/Sr1-O1 | 2.896(2) | 2.904(3) |
| 1 × La1/Sr1-O2a | 2.358(2) | 2.354(3) |
| 1 × La1/Sr1-O2b | 2.443(17) | 2.443(20) |
| 3 × Sr2-O1 | 2.496(3) | 2.480(3) |
| 3 × Sr2-O1 | 2.620(3) | 2.628(3) |
| 1 × Sr2-O2a$^a$ | 2.772(7) | 2.779(8) |
| 1 × Sr2-O2b$^a$ | 2.407(56) | 2.404(64) |

$^a$ only shortest distance to oxygen ion occupying six-fold degenerate site 18$g$ (O2a)/18$h$ (O2b)

dynamic from static disorder, refinement calculations were carried out on neutron diffraction data collected at 1.5 K. Since the neutron scattering lengths of lanthanum and strontium are similar (8.24 and 7.02 fm, respectively), the La1/Sr1 ratio was fixed to the SOFs derived from the XRD data at room temperature. The refinement results show that the anisotropic displacement parameters for the O1 site are comparable to those obtained from joint refinement at room temperature, favoring a static disorder of these ions.

Structure refinement of $La_{1.2}Sr_{2.7}IrO_{7+\delta}$

The structure of the new compound $La_{1.2}Sr_{2.7}IrO_{7+\delta}$ was refined using simultaneously X-ray and neutron diffraction data in analogy to the ruthenate. $La_{1.2}Sr_{2.7}IrO_{7+\delta}$ also crystallizes in a rhombohedral structure (space group $R$-3$m$) with lattice constants $a$ = 5.771(1) Å and $c$ = 18.348(3) Å. As a starting model for the structural refinement, atomic coordinates of the ruthenate were used. The joint refinement of the data sets converged smoothly and yielded atomic parameters and selected bond lengths listed in Table 1 and 2, respectively. The observed and calculated X-ray and neutron powder diffraction patterns are shown in Figure 2 (bottom). The total oxygen content is 7.35(3) favoring the above-discussed model with oxide and peroxide ions at the O2 site, in accordance with the ruthenate. Heating the sample to 1200°C resulted in the formation of iridium metal, $La_2O_3$, SrO and water, which evaporates. From the obtained weight loss of 6.83%, a total oxygen content of 7.41 is deduced. Taking into account the experimental errors, this value is very similar to the results gained from the Rietveld refinement.

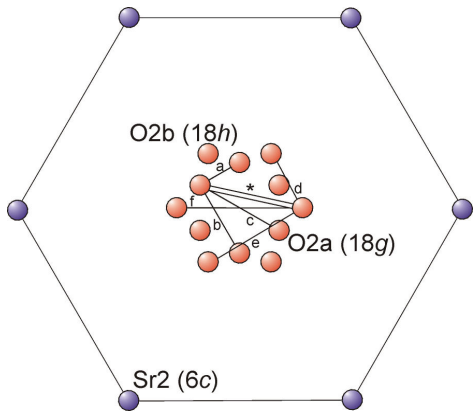

**Fig. 3** View of the (A'$_2$O$_{1+\delta}$) layer of $La_{1.2}Sr_{2.7}MO_{7+\delta}$ perpendicular to the $c$-axis. Isotropic displacement parameters are drawn at the 50% probability level. Distances (in Å) between oxygen ions of the same site are (a) 0.665/0.668, (b) 1.151/1.157, (c) 1.329/1.336, (d) 0.921/0.934, (e) 1.595/1.618, and (f) 1.842/1.868 for $M$ = Ru/Ir. The asterisk indicates the position of one peroxide unit with a bond length of 1.533 and 1.549 Å for $La_{1.2}Sr_{2.7}RuO_{7+\delta}$ and $La_{1.2}Sr_{2.7}IrO_{7+\delta}$, respectively.

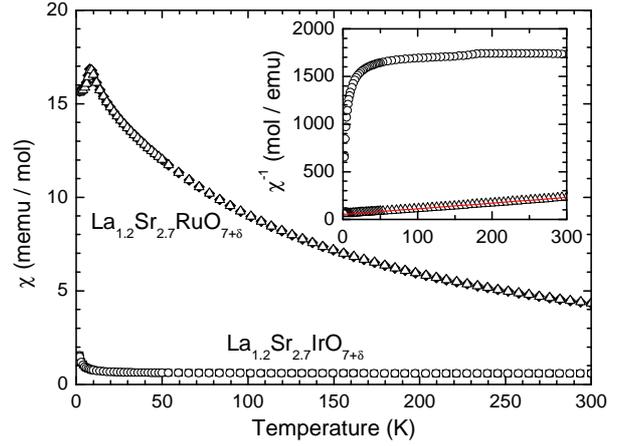

**Fig. 5** Temperature dependence of ZFC (open symbols) and FC (closed symbols) molar magnetic susceptibilities measured in an applied field of 20 kG for $La_{1.2}Sr_{2.7}IrO_{7+\delta}$ and $La_{1.2}Sr_{2.7}RuO_{7+\delta}$. In the inset the inverse susceptibilities are shown. For the ruthenate the fitted Curie-Weiss law is displayed.

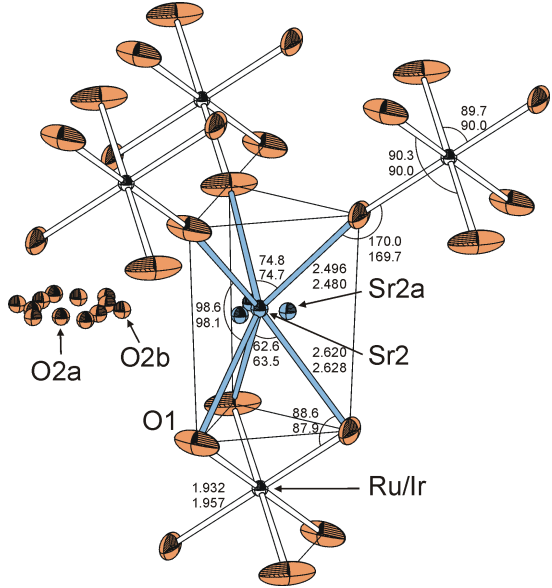

**Fig. 4** Detailed view of the crystal structure showing the connection between triangular $SrO_6$ prisms and $MO_6$ octahedra as well as selected bond lengths (Å) in $La_{1.2}Sr_{2.7}MO_{7+\delta}$ with $M$ being Ru (upper numbers) and Ir (lower numbers), respectively. Displacement ellipsoids are drawn at the 50% probability level.

*Comparison of the structures*

For both compounds, the $BO_6$ octahedra maintain an almost perfect $O_h$ symmetry with O-$M$-O bond-angles close to 90°. The experimentally found B-O bond lengths are 1.932(2) and 1.957(3) Å for the ruthenate and iridate, respectively. The electron density of the O1 positions is modeled as an ellipsoid (see Figure 4). This leads to an artificial bond shortening, i.e. the determined atomic distances are shorter than the real bond lengths. Nevertheless, the experimental results can be compared with the values deduced from bond valence sum calculations using the method of Brown and Altermatt.[18] Since magnetic and XANES results (*vide infra*) yield an oxidation state of +5 for the transitions metal cations, the bond valence parameters 1.888 Å[19] and 1.916 Å[20] for $Ru^{5+}$ and $Ir^{5+}$, respectively, were used in the calculations. For the Ru-O bond length a value of 1.955 Å is expected, whereas the distance in the iridate is calculated to be 1.983 Å. These distances are about 0.02 Å longer than the experimentally determined ones, a finding consistent with the above-mentioned artificial bond shortening.

**Magnetic measurements**

Recent magnetic investigations on $La_{1.2}Sr_{2.7}RuO_{7+\delta}$ indicated the presence of an antiferromagnetic interaction between the ruthenium centers associated with the occurrence of a long-range antiferromagnetic ordering at low temperatures. Owing to the close structural similarity, one would also expect antiferromagnetic coupling for the iridate. A comparison of the magnetic susceptibilities of $La_{1.2}Sr_{2.7}RuO_{7+\delta}$ and $La_{1.2}Sr_{2.7}IrO_{7+\delta}$ in an applied field of 20 kG is shown in Figure 5. The data for the ruthenate reveal an antiferromagnetic transition at 7 K consistent with previous results, whereas the susceptibility of the iridate proves that this compound is paramagnetic down to 1.8 K without any magnetic ordering. For both compounds, the ZFC and FC data completely overlay at all temperatures except for a minimal deviation at the very lowest temperatures measured. The inverse susceptibility of $La_{1.2}Sr_{2.7}IrO_{7+\delta}$ (inset of Figure 5) clearly demonstrates that unlike the ruthenate, no Curie-Weiss behavior is observed. The explanation for this unusual magnetic behavior of $La_{1.2}Sr_{2.7}IrO_{7+\delta}$ results from the strong spin-orbit coupling observed in the 5$d$ transition metals.[21] From the XANES measurements discussed below, the valence of Ir was found to be +5. Due to the strong crystal field splitting in the 5$d$ series, the $Ir^{5+}$ ions have a low-spin configuration ($t_{2g}^4 e_g^0$). The resulting ground term in an octahedral crystal field is $^3T_{1g}$. For $T$ terms a reduced orbital angular momentum of L' = 1 remains, which couples with the spin of S = 1 giving rise to three energy eigenstates with J = 0,1, and 2. According to Kotani's theory on the magnetic moment of complex ions,[22] the effective magnetic moment depends on the thermal population of these states, which is expressed as the ratio of temperature ($k_BT$) to spin-orbital coupling constant ($\zeta$) as follows:

$$n_{\text{eff}}^2 = \frac{3}{2} y \cdot \frac{24 + (1/y - 9)\exp(-1/y) + (5/y - 15)\exp(-3/y)}{1 + 3\exp(-1/y) + 5\exp(-3/y)}$$

with $\quad y \equiv \dfrac{2k_BT}{\zeta}$.

This behavior is displayed in the inset of Figure 6. For $Ir^{5+}$ the spin-orbit coupling is strong, i.e. $k_BT << \zeta$ ($y << 1$). In this case the square of the effective magneton number $n_{\text{eff}}^2$ is proportional to the temperature $T$ (Figure 6) according to

$$n_{\text{eff}}^2 \approx 36y = 72 \cdot \frac{k_BT}{\zeta}.$$

Since the magnetic susceptibility is given by

$$\chi_{\text{mole}} = \frac{N_L^2 \mu_0 \mu_B^2 n_{\text{eff}}^2}{3RT},$$

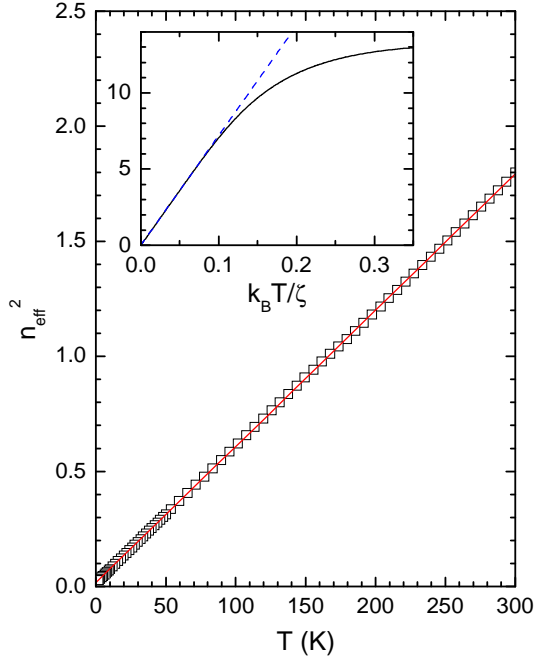

**Fig. 6** Temperature dependence of the square of the effective magneton number for $La_{1.2}Sr_{2.7}IrO_{7+\delta}$ and the corresponding linear fit. The inset shows $n_{eff}^2$ vs. $k_BT/\zeta$ according to Kotani's formula for $^3T_1$ terms and the linear behavior (dashed line) for $k_BT \ll \zeta$.

it follows that the magnetic susceptibility $\chi$ is independent of $T$. The experimentally observed dependence of $n_{eff}^2$ on temperature for $La_{1.2}Sr_{2.7}IrO_{7+\delta}$ is illustrated in Figure 6. Data were fitted with a linear regression according to

$$n_{eff}^2 = A + B \cdot T$$

with $A = 0.0159(3)$ and $B = 0.005921(2)$. The very small value $A$ strongly indicates that nearly all iridium ions are in the oxidation state +5, based on the fact that the magnetic susceptibility for $Ir^{6+}$ (ground term $^4A_{2g}$) and $Ir^{4+}$ (ground term $^2T_{2g}$) ions would follow a Curie-Weiss law. The spin-orbital coupling constant $\zeta$ can be calculated from the equations above to be ~ 8450 cm$^{-1}$. This value is in good agreement with results reported in earlier publications,[21,23,24] where $\zeta$ is in the range between 5000 and 9000 cm$^{-1}$.

**XANES measurements**

The above-mentioned results already show that the oxidation state of iridium in $La_{1.2}Sr_{2.7}IrO_{7+\delta}$ is +5. Nevertheless, we were looking for a more direct method to investigate the electronic structure of the transition metal. For this reason, XANES studies have been systematically carried out at the Ir-L$_{III}$ edge for a series of compounds, namely $La_2ZnIrO_6$, $IrO_2$, $Sr_2IrO_4$, $Sr_2FeIrO_6$, $Sr_2YIrO_6$, and $Sr_2CaIrO_6$. In these oxides the iridium ions are stabilized in an octahedral oxygen environment with the oxidation states +4 ($d^5$), +5 ($d^4$), and +6 ($d^3$). For $La_{1.2}Sr_{2.7}RuO_{7+\delta}$ similar measurements at the ruthenium L$_{III}$–absorption edge were carried out. The results proved that Ru is in the oxidation state +5. Details of these investigations can be found in Ref.13.

Figure 7 presents normalized Ir-L$_{III}$ edge XANES spectra for the studied perovskite and the reference compounds. For all samples a pronounced white line corresponding to transitions of 2$p$ core electrons to unoccupied bound states with predominantly $d$ character can be observed. It has been generally accepted to analyze these edge spectral features by fitting them with a combination of two pseudo-Voigt and one arctangent functions. These functions represent transitions to bound states and to the continuum, respectively. To improve the resolution, the first or second derivatives of the spectra are usually used. If the second derivative is fitted, the contribution of the arctangent becomes so small that it can well be neglected.[25-27] For ruthenium containing perovskite oxides, there is a clear splitting of the white line into two peaks assigned to transitions to $t_{2g}$ and $e_g$ states.[28-30] Due to the higher natural line width[31] and the lower instrumental resolution at higher energies, the iridium L$_{III}$ XANES

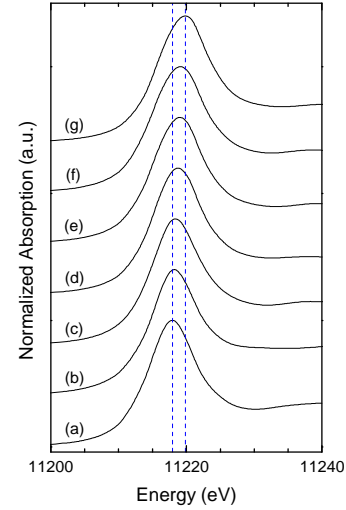

**Fig. 7** Normalized Ir-L$_{III}$ edge XANES spectra for (a) $La_2ZnIr^{IV}O_6$, (b) $Ir^{IV}O_2$, (c) $Sr_2Ir^{IV}O_4$, (d) $Sr_2FeIr^{V}O_6$, (e) $Sr_2YIr^{V}O_6$, (f) $La_{1.2}Sr_{2.7}IrO_{7+\delta}$, and (g) $Sr_2CaIr^{VI}O_6$. The two dashed lines illustrate the white line shift to higher energies with increasing oxidation state of iridium.

spectra exhibit only one broad feature rather than two independent peaks. In order to estimate peak positions more accurately, curve fitting was done using the first-derivative spectrum. The fit for the title compound $La_{1.2}Sr_{2.7}IrO_{7+\delta}$ is represented in Figure 8 (left). The obtained energy positions of both pseudo-Voigt functions are plotted vs. the oxidation state of iridium in Figure 8 (right). For the reference samples $La_2ZnIr^{IV}O_6$, $Ir^{IV}O_2$, $Sr_2Ir^{IV}O_4$, $Sr_2FeIr^{V}O_6$, $Sr_2YIr^{V}O_6$, and $Sr_2CaIr^{VI}O_6$, a linear increase to higher energies with increasing oxidation state of iridium can be observed. It should be mentioned that for $IrO_2$ and $Sr_2IrO_4$ it was only possible to fit one instead of two pseudo-Voigt functions. The obtained energy

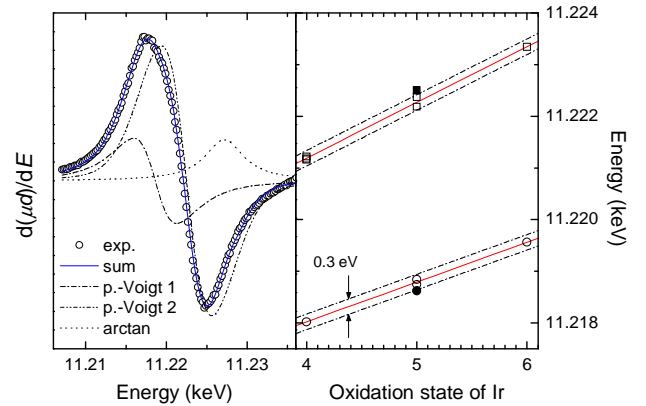

**Fig. 8** Left: Fit of the first derivative of the Ir-L$_{III}$ XANES data for $La_{1.2}Sr_{2.7}IrO_{7+\delta}$ using two pseudo-Voigt and one arctangent functions. Right: Plots of the energy position corresponding to transitions to $t_{2g}$ (circles) and $e_g$ (squares) states vs. the oxidation state of iridium in the studied compounds. Solid lines represent linear fits for the reference materials, solid symbols represent the energy values for $La_{1.2}Sr_{2.7}IrO_{7+\delta}$

values for $La_{1.2}Sr_{2.7}IrO_{7+\delta}$ are within an estimated accuracy of ± 0.2 in good agreement with the oxidation state of +5. The average energy splitting between transitions to $t_{2g}$ and $e_g$ states increases from 3.17 eV for $Ir^{4+}$ to 3.79 eV for $Ir^{6+}$, consistent with previous XANES studies on iridium containing perovskites.[26,27] The comparatively large energy splitting of 3.88 eV for $La_{1.2}Sr_{2.7}IrO_{7+\delta}$ can be explained by significant changes in the covalent character of the Ir-O bonds.[27] In analogy to the investigated double perovskites $A_2A'IrO_6$ (A' = La, Sr; A = Zn, Fe, Y, Ca), the Ir-O bond in the studied compound $(A'_2O_{1+\delta})(A_2IrO_6)$ (A = La/Sr; A' = Sr) competes with the A'-O and A-O bonds by sharing the same oxygen 2p orbitals via a pathway of nearly 180° and 90°, respectively. The degree of covalency of the Ir-O bond is therefore altered by the ionicity/covalency of these competing bonds. For the latter, Pauling's ionicities, which are deduced from the electronegativities of the participating atoms, can be used as a measure. Comparing for example $Sr_2FeIrO_6$ and $La_{1.2}Sr_{2.7}IrO_{7+\delta}$, ionicities for the Fe-O and La/Sr-O bonds are 0.502 and 0.763/0.803, respectively. Due to the higher average ionicity of the A-O and A'-O bonds in $La_{1.2}Sr_{2.7}IrO_{7+\delta}$, the Ir-O bond covalency in the title compound is higher than that in $Sr_2FeIrO_6$. This enhanced Ir-O bond covalency in turn leads to a larger energy splitting of the $t_{2g}$ and $e_g$ states consistent with our results.

**Conclusion**

Solid state synthesis was used to prepare the new compound, $La_{1.2}Sr_{2.7}IrO_{7+\delta}$, which is the third example for an $n = 2$ member of the $(A'_2O_{1+\delta})(A_nB_{n-1}O_{3n})$ family of hexagonal perovskite related oxides. Both title compounds, $La_{1.2}Sr_{2.7}RuO_{7+\delta}$ and $La_{1.2}Sr_{2.7}IrO_{7+\delta}$, have the transition metal cations in an unusually high oxidation state as confirmed by the temperature dependence of the magnetic susceptibility as well as by XANES investigations. In contrast to the ruthenate, which shows a Curie-Weiss behavior with an antiferromagnetic transition at 7 K, the magnetic susceptibility of the iridate is almost temperature-independent and can be explained by a large spin-orbit coupling typical for 5d transition metals.

With regard to the oxygen stoichiometry, a $\delta$ value of about 1/3 for both the ruthenate and the iridate was determined by joint Rietveld refinement of X-ray and neutron diffraction data. This finding strongly confirms a structural model with both oxide and peroxide ions within the $(A'_2O_{1+\delta})$ layers. The coexistence of $O^{2-}$ and $(O_2)^{2-}$ is also supported by the determined O-O bond length of 1.533 Å, which is typical for peroxide ions. To increase the $\delta$ value to the upper limit of 1, which corresponds to a compound with all oxide ions in the $(A'_2O)$ layers substituted for peroxide units, high-pressure syntheses will be carried out in the near future.

**Acknowledgements**

This work is based on experiments performed at the Swiss spallation neutron source SINQ, Paul Scherrer Institute, Villigen, Switzerland. We also gratefully acknowledge the provision of beamtime and resources at HASYLAB. Financial support was received from the Deutsche Forschungsgemeinschaft within SFB 484 and from the EU via access programs.